%% file: main.tex
\documentclass[sigconf]{acmart}

\usepackage{listings}
\usepackage{enumitem} 
\usepackage{multicol}

\usepackage{longtable}

\usepackage{longtable}

\AtBeginDocument{%
  \providecommand\BibTeX{{%
    \normalfont B\kern-0.5em{\scshape i\kern-0.25em b}\kern-0.8em\TeX}}}

\copyrightyear{2025}
\acmYear{2025}
\setcopyright{rightsretained}
\acmConference[SIGCSE TS 2025]{Proceedings of the 56th ACM Technical Symposium on Computer Science Education V. 2}{February 26-March 1, 2025}{Pittsburgh, PA, USA}
\acmBooktitle{Proceedings of the 56th ACM Technical Symposium on Computer Science Education V. 2 (SIGCSE TS 2025), February 26-March 1, 2025, Pittsburgh, PA, USA}
\acmDOI{10.1145/3641555.3705227}
\acmISBN{979-8-4007-0532-8/25/02}

\begin{document}
\title[The Effects of LLM-Powered Personalized Parsons Puzzle]{Personalized Parsons Puzzles as Scaffolding Enhance Practice Engagement Over Just Showing LLM-Powered Solutions}

\author{Xinying Hou}
\orcid{0000-0002-1182-5839}
\affiliation{%
  \institution{University of Michigan}
  \city{Ann Arbor}
  \state{Michigan}
  \country{USA}
}
\email{xyhou@umich.edu}

\author{Zihan Wu}
\orcid{0000-0002-3161-2232}
\affiliation{%
  \institution{University of Michigan}
  \city{Ann Arbor}
  \state{Michigan}
  \country{USA}
}
\email{ziwu@umich.edu}

\author{Xu Wang}
\orcid{}
\affiliation{%
  \institution{University of Michigan}
  \city{Ann Arbor}
  \state{Michigan}
  \country{USA}
}
\email{xwanghci@umich.edu}

\author{Barbara J. Ericson}
\orcid{0000-0001-6881-8341}
\affiliation{%
  \institution{University of Michigan}
  \city{Ann Arbor}
  \state{Michigan}
  \country{USA}
}
\email{barbarer@umich.edu}






\begin{abstract}
As generative AI products could generate code and assist students with programming learning seamlessly, integrating AI into programming education contexts has driven much attention. However, one emerging concern is that students might get answers without learning from the LLM-generated content. In this work, we deployed the LLM-powered personalized Parsons puzzles as scaffolding to write-code practice in a Python learning classroom (PC condition) and conducted an 80-minute randomized between-subjects study. Both conditions received the same practice problems. The only difference was that when requesting help, the control condition showed students a complete solution (CC condition), simulating the most traditional LLM output. Results indicated that students who received personalized Parsons puzzles as scaffolding engaged in practicing significantly longer than those who received complete solutions when struggling. 
\end{abstract}
\vspace{-12mm}



\maketitle

\input{Sections/1_introduction}
\input{Sections/3_Methods}

\input{Sections/4_Results}

\input{Sections/5_Discussion_FutureWork}
\bibliographystyle{ACM-Reference-Format}
\bibliography{reference}


\end{document}

%% file: Sections/1_introduction.tex
\section{Introduction}
Writing code is challenging for most novices. Thanks to the development of large language models (LLMs) and generative AI techniques, students now have the opportunity to generate code directly based on a natural language description \cite{kazemitabaar2023novices}. Therefore, some novices can simply use AI code generation tools to complete their short programming task homework without even engaging with the programming practice \cite{kazemitabaar2023novices}. This raises growing concerns about over-utilizing generative AI tools when learning programming, which can harm students' programming skill development. 

To support struggling students in programming practice while keeping them engaged, built upon the previous work \cite{hou2024codetailor}, this work explored the use of personalized Parsons puzzles as write-code scaffolding in a real classroom lecture setting. Parsons puzzles are an increasingly popular active programming exercise that requires students to arrange a set of drag-and-drop code blocks to solve a problem. It can have distractor blocks that are not needed in a correct solution. The personalized Parsons puzzle applied as programming scaffolding applied two levels of personalization when offered as scaffolding \cite{hou2022using}. The delivered puzzle was tailored to students' existing code at both the code solution level and the block level \cite{hou2024codetailor}. To understand the effectiveness of this scaffolding technique, we deployed it to a real undergraduate classroom setting and conducted a randomized between-subjects classroom experiment. The control condition offered students a complete AI-generated code solution when requesting help, a typical output of most AI code-generation tools. Students in both conditions could choose to ask for the provided scaffolding or solve the write-code practice independently. This work presents the preliminary results addressing the key RQ: \textit{Are there condition differences in terms of students' practice engagement?}


%% file: Sections/3_Methods.tex
\vspace{-2mm}
\section{Methods}
The study was conducted in the winter semester of 2024 at a large public research university in the northern United States. This course covered programming concepts including Python basic data structures, object-oriented programming concepts, and more. 

\textbf{Two study conditions} Students in both conditions practiced traditional short programming tasks in an online system. Each task contained a natural language description, a programming area, and a "Save \& Run" button to execute the code and display the unit test results. It also included a "Help" button to trigger the support. Students could ask for support at any time or choose to solve the write-code problems independently. In the puzzle-scaffolding condition (PC), students received personalized Parsons puzzles as scaffolding. After clicking the "Help" button, PC students received a personalized Parsons puzzle with no indentation-level requirements and immediate feedback. Upon completing a puzzle, they could copy the solution to the clipboard with the "Copy Answer to Clipboard" button or retyping it. Students could also regenerate a new personalized Parsons puzzle at any point. Students in the control condition (CC) practiced programming with the same programming interface. However, after clicking "Help", they received a complete personalized correct code solution. They could also copy the solution to the clipboard with the "Copy Answer to Clipboard" button. Both two types of support conducted the same personalization on the solution level \cite{hou2024codetailor}. This study used the institution-protected GPT-4, which protected data appropriately. 
\vspace{-3mm}
\begin{figure}[ht]
    \centering
    \includegraphics[width=0.9\linewidth]{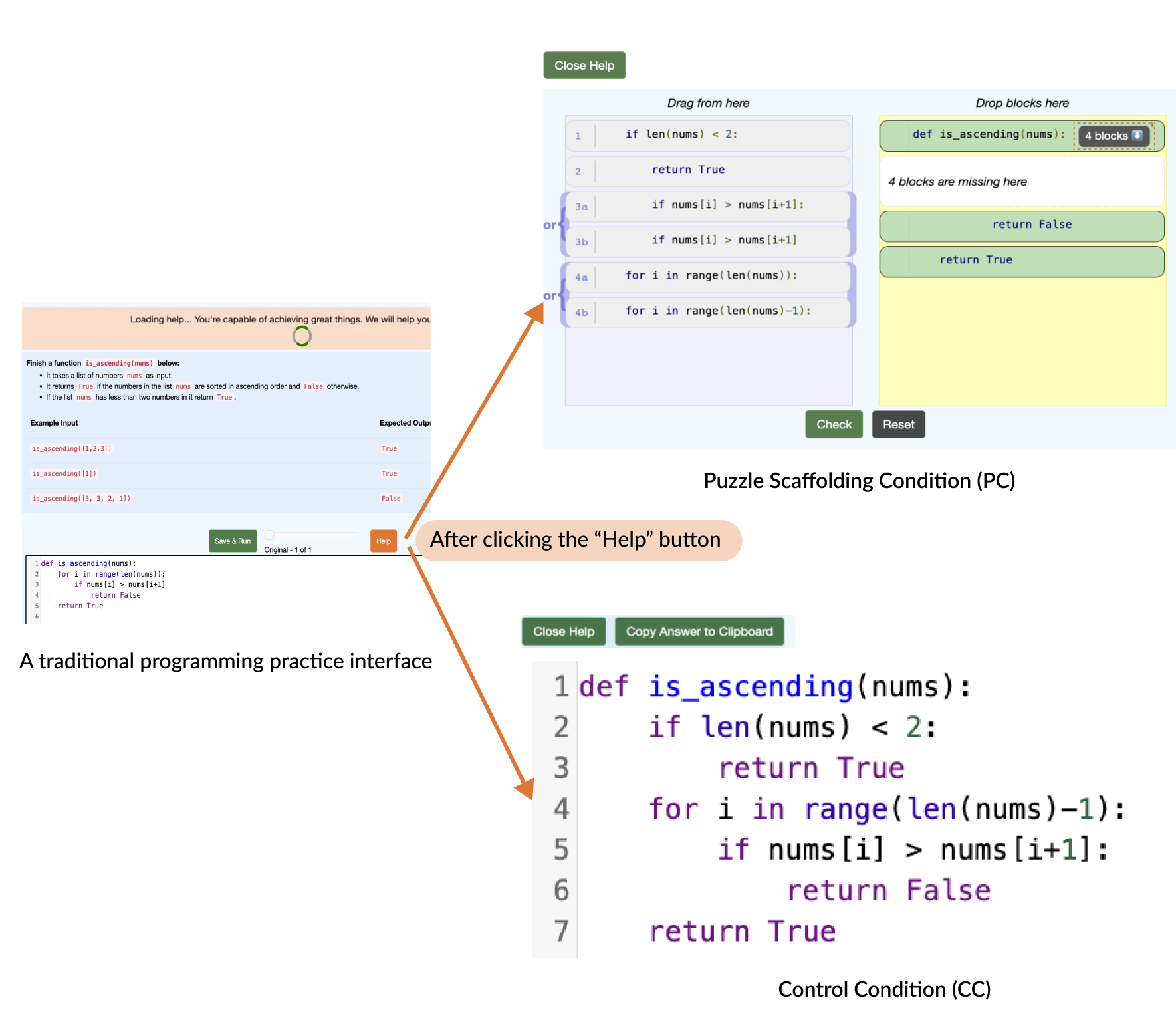}
    \caption{Two conditions: Puzzle Scaffolding condition (PC) and Control Condition (CC)}
    \label{interface}
    \vspace{-5mm}
\end{figure}



\textbf{Participants and Procedure} This one-session classroom study was conducted during an 80-minute lecture period in week six. This study was introduced to students as an in-class practice. Students were instructed to do the sections in order and answer questions to the best of their ability without outside help. Students were randomly assigned to one of two support conditions. The practice topic was \textit{nested dictionaries in Python}. Students were given four write-code practice problems in each condition. Each section had no time restrictions, allowing students to progress through the materials at their own pace until the end of the lecture. 


%% file: Sections/4_Results.tex
\vspace{-3mm}
\section{Results}
There were 118 students (51 PC students and 67 CC students) who worked on all four practice questions as instructed and did not encounter any technical difficulties. We used the Mann–Whitney U test instead of the t-test in cases where the data was not normally distributed. Results showed that students in both conditions had similar levels of pretest performance (\textit{U} = 1628.5, \textit{p} = .649, CLES = .48) and self-efficacy levels (\textit{U} = 1595.5, \textit{p} = .540, CLES = .47) before the study, suggesting that the conditions were comparable.

To answer the RQ, we investigated students' practice time (in minutes), which indicated the duration of the practice engagement. Students' practice time was calculated as the sum of the time they spent on each practice question. Results showed that students in the PC condition (\textit{M} = 22.7, \textit{SD} = 10.1, \textit{Median} = 22.8) spent significantly more time with the practice than those in the CC condition (\textit{M} = 15.8, \textit{SD} = 12.7, \textit{Median} = 11.7), \textit{U} = 2368.0, \textit{p} < .001, CLES = .69. Specifically, PC students spent approximately 7 more minutes on average in the practice compared to students in the CC condition. On the contrary, six students in the CC condition finished practice in less than two minutes. Some mainly opened the programming help immediately after receiving the question, copied the solution to the write-code box, and immediately submitted it. 

However, according to the students' perceptions, we found some PC students felt that the puzzles still provided too much support, and they would rather have a more lightweight approach, like subtle hints. However, another group of PC students required more guidance on how the blocks should be rearranged. They asked for additional help when they did not know where to place the code blocks. For CC students, some were satisfied with providing a solution as a quick and accurate way to deliver support, but some found providing a full solution prevented them from thinking independently and learning anything besides getting the right answer. Such conflicted opinions uncovered the trade-off regarding the question \textit{"when to provide what types of support"} in the large learning settings.

%% file: Sections/5_Discussion_FutureWork.tex
\vspace{-2mm}
\section{Limitations \& Future work \& Conclusion}
\textbf{Limitations} Because the course time was only 80 minutes, students in the PS condition spent more time practicing, leaving insufficient time for the designed posttest. This prevented us from exploring how students perform differently after practicing with the two types of support. In addition, we encountered technical difficulties for some students, which decreased their learning experience.

\textbf{Next Steps} We plan to look into other metrics to investigate students' practice engagement from their behavior log data, such as their practice attempts. In addition, the current design allows students to use the help feature at any point they want, even before they write any code. We plan to explore developing an optimized support trigger in the future.

\textbf{Conclusion} As large language models and AI code-generation tools become more prevalent, educators' concerns about the over-utilization of these generation tools leading to fake practice progress and cheating are growing. Our results showed positive evidence that providing a personalized Parsons puzzle as programming support can significantly engage students with practice problems more than showing them a complete correct solution. By providing students with an active learning activity when requested scaffolding instead of a passive textual answer, students engaged longer with the materials and invested more effort, which potentially led to deeper cognitive engagement with the practice question. 

\vspace{-1mm}